\documentclass[aps,english,showpacs,twocolumn]{revtex4-1}
\usepackage{amssymb}
\usepackage{amsmath}
\usepackage{graphicx}
\usepackage{epsfig}

\begin{document}

\title{Partitioning technique for a discrete quantum system}
\author{L. Jin and Z. Song}
\email{songtc@nankai.edu.cn}
\affiliation{School of Physics, Nankai University, Tianjin 300071, China}

\begin{abstract}
We develop the partitioning technique for quantum discrete systems. The
graph consists of several subgraphs: a central graph and several branch
graphs, with each branch graph being rooted by an individual node on the
central one. We show that the effective Hamiltonian on the central graph can
be constructed by adding additional potentials on the branch-root nodes,
which generates the same result as does the the original Hamiltonian on the
entire graph. Exactly solvable models are presented to demonstrate the main
points of this paper.
\end{abstract}

\maketitle

%03.65.-w Quantum mechanics
%03.65.Nk Scattering theory
%11.30.Er Charge conjugation, parity, time reversal, and other discrete symmetries

\section{Introduction}

The Schr\"{o}dinger equation lies at the heart of quantum mechanics. Secular
equation has analytic solutions only for a few very special cases.
Approximation techniques and computational methods have been developed for
treating such problem. Many of them are rooted in the partitioning technique
\cite{Lindgren,Hubac} which was introduced by Feshbach \cite{Feshbach} and L%
\"{o}wdin \cite{Lowdin} independently. Discrete models, including quantum
networks, have been a cornerstone of theoretical explorations due to their
analytical and numerical tractability \cite{WXG}, the availability of exact
solutions, and the ability to capture counter-intuitive physical phenomena,
such as non-spreading wavepacket \cite{Kim} and Bloch oscillation \cite%
{FBloch,CZener,Hartmann} in linear chain. In recent years, optical lattice
\cite{IBloch,Zoller}, photonic crystal \cite{KML,UG}, etc. have increasingly
permitted the experimental exploration of quantum discrete models.

In this paper, we study the partitioning technique for quantum discrete
systems. The concerned graph consists of several subgraphs: a central graph
and several branch graphs, with each branch graph being rooted by an
individual node on the central one. Applying the projection theory \cite%
{Lowdin} to such a graph, we show that the effective Hamiltonian on the
central graph can be constructed by adding additional potentials on the
branch-root nodes, which generates the same result as does the the original
Hamiltonian on the entire graph. As the demonstration, we present two
exactly solvable models, which correspond to real and imaginary potentials.

This paper is organized as follows. Section \ref{sec_Partition_tech} shows a
formalism for the partitioning technique in discrete quantum systems.
Section \ref{sec_central_graph} is the heart of this paper which presents a
method to obtain the projection Hamiltonian. Section \ref{sec_illus}
consists of two exactly solvable examples to illustrate our main idea.
Section \ref{sec_conclusion} is the summary and discussion.

\section{Partitioning technique}

\label{sec_Partition_tech}
%%%%%%%%%%%%%%%%%%%%%%%%%%%%%%%%%%%%%%%%%%%%%%%%%%%%%%%%%%%%%%%%%%%%%%%%

\begin{figure}[tbp]
\includegraphics[ bb=25 20 569 478, width=6.8 cm, clip]{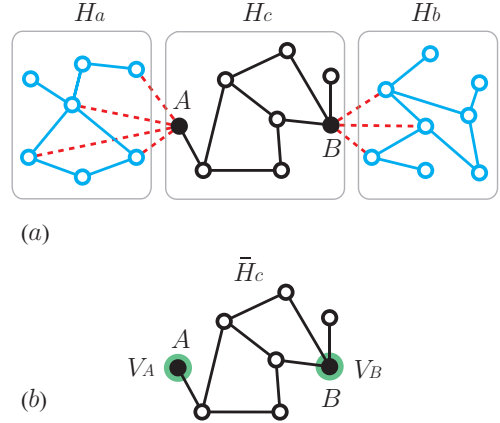}
\caption{(Color online)(a) Schematic illustration of the graph
consisted of two branch graphs $a$, $b$ (cyan) and a center graph
$c$ (black). The dashed edges (red) represent the connections
between them with $A$ and $B$ being the branch-root nodes. (b) The
L\"{o}wdin's projection Hamiltonian for the center graph which is
composed of the original Hamiltonian $H_{c}$ and additional on-site
potentials $V_{A}$ and $V_{B}$ on the nodes $A$ and $B$,
respectively.} \label{illus}
\end{figure}

%%%%%%%%%%%%%%%%%%%%%%%%%%%%%%%%%%%%%%%%%%%%%%%%%%%%%%%%%%%%%%%%%%%%%%%%

L\"{o}wdin has developed a partitioning technique in the algebra of
matrices, with which various self-consistent field methods can be nicely
formulated. In this procedure, the original Hamiltonian is simply
transformed in a chosen discrete representation. The entire space is usually
divided into two subspaces, named a model space and an orthogonal space. The
basic idea is to find an effective Hamiltonian which acts only within the
target model space but generates the same result as the original Hamiltonian
acting on the complete space \cite{Lindgren,Hubac}. The partitioning
technique enables us focus our interest on certain part of the system. In
general, the effective Hamiltonian cannot be obtained explicitly, but
provides a formalism to develop perturbation method.

In the following we will show that, when the technique is applied to a
specific discrete system, the effective Hamiltonian is of realistic
significance. We consider a quantum graph, which is a collection of nodes
and edges. It is also equivalent to a single-particle tight-binding model.
For simplicity, we partition the complete graph into three subgraphs, a
central part $c$, two independent branches $a$ and $b$.

The Hamiltonian (or connectivity matrix) of such a graph has the form

\begin{equation}
H=\left(
\begin{array}{ccc}
H_{a} & H_{ac} & 0 \\
H_{ca} & H_{c} & H_{cb} \\
0 & H_{bc} & H_{b}%
\end{array}%
\right) ,  \label{H}
\end{equation}%
where

\begin{eqnarray}
H_{a} &=&-\sum_{i,j=1}^{N_{a}}\left( \kappa _{ij}^{a}\left\vert
i\right\rangle _{a}\left\langle j\right\vert +\text{H.c.}\right) ,
\label{H_a} \\
H_{b} &=&-\sum_{i,j=1}^{N_{b}}\left( \kappa _{ij}^{b}\left\vert
i\right\rangle _{b}\left\langle j\right\vert +\text{H.c.}\right) ,
\label{H_b} \\
H_{c} &=&-\sum_{i,j=1}^{N_{c}}\left( \kappa _{ij}^{c}\left\vert
i\right\rangle _{c}\left\langle j\right\vert +\text{H.c.}\right) ,
\label{H_c}
\end{eqnarray}%
Here $N_{\gamma =a,b,c}$ denotes the dimension of the three subgraphs. $%
\kappa _{ij}^{\gamma =a,b,c}$ denotes the coupling between $i$ and $j$ of
the graph $\gamma $, and reduces to the on-site potential for $i=j$. The
connections between the subgraphs are

\begin{eqnarray}
H_{ca} &=&H_{ac}^{\dagger
}=-\sum_{i=1}^{N_{c}}\sum_{j=1}^{N_{a}}g_{ij}^{a}\left\vert i\right\rangle
_{ca}\left\langle j\right\vert , \\
H_{cb} &=&H_{bc}^{\dagger
}=-\sum_{i=1}^{N_{c}}\sum_{j=1}^{N_{b}}g_{ij}^{b}\left\vert i\right\rangle
_{cb}\left\langle j\right\vert .
\end{eqnarray}%
where $g_{ij}^{\gamma =a,b}$ is the coupling strength between $\left\vert
j\right\rangle _{\gamma }$ and branch-root nodes $\left\vert i\right\rangle
_{c}$.

Our aim is the solution of the Schr\"{o}dinger equation

\begin{equation}
H\left\vert f_{k}\right\rangle =E_{k}\left\vert f_{k}\right\rangle ,
\label{S_eq}
\end{equation}%
where

\begin{equation}
\left\vert f_{k}\right\rangle =\sum_{\gamma =a,b,c}\sum_{l=1}^{N_{\gamma
}}f_{k}^{\gamma }\left( l\right) \left\vert l\right\rangle _{\gamma }.
\label{fk}
\end{equation}%
Then the Schr\"{o}dinger equation can be written in the matrix form

\begin{equation}
\left(
\begin{array}{ccc}
H_{a} & H_{ac} & 0 \\
H_{ca} & H_{c} & H_{cb} \\
0 & H_{bc} & H_{b}%
\end{array}%
\right) \left(
\begin{array}{c}
f_{k}^{a} \\
f_{k}^{c} \\
f_{k}^{b}%
\end{array}%
\right) =E_{k}\left(
\begin{array}{c}
f_{k}^{a} \\
f_{k}^{c} \\
f_{k}^{b}%
\end{array}%
\right) ,  \label{matrix}
\end{equation}%
and more explicit form%
\begin{eqnarray}
H_{a}f_{k}^{a}+H_{ac}f_{k}^{c} &=&E_{k}f_{k}^{a},
\label{General_Schrodinger_Eq_a} \\
H_{c}f_{k}^{c}+H_{ca}f_{k}^{a}+H_{cb}f_{k}^{b} &=&E_{k}f_{k}^{c},
\label{General_Schrodinger_Eq_c} \\
H_{b}f_{k}^{b}+H_{bc}f_{k}^{c} &=&E_{k}f_{k}^{b}.
\label{General_Schrodinger_Eq_b}
\end{eqnarray}%
Under the condition of the existence of the inverse matrices $\left(
E_{k}-H_{a}\right) ^{-1}$ and $\left( E_{k}-H_{b}\right) ^{-1}$, we have

\begin{eqnarray}
f_{k}^{a} &=&\left( E_{k}-H_{a}\right) ^{-1}H_{ac}f_{k}^{c},  \label{fk_a} \\
f_{k}^{b} &=&\left( E_{k}-H_{b}\right) ^{-1}H_{bc}f_{k}^{c},  \label{fk_b}
\end{eqnarray}%
Then the L\"{o}wdin's projection Hamiltonian $\bar{H}_{c}$ has the form

\begin{equation}
\bar{H}_{c}=H_{c}+\bar{H}_{a}+\bar{H}_{b},  \label{H_eff_c}
\end{equation}%
where

\begin{eqnarray}
\bar{H}_{a} &=&H_{ca}\left( E_{k}-H_{a}\right) ^{-1}H_{ac},  \label{H_a_eff}
\\
\bar{H}_{b} &=&H_{cb}\left( E_{k}-H_{b}\right) ^{-1}H_{bc}.  \label{H_b_eff}
\end{eqnarray}%
Remarkably, the corresponding Schr\"{o}dinger equation for the subgraph $c$
(Eq. (\ref{General_Schrodinger_Eq_c})) is reduced to

\begin{equation}
\bar{H}_{c}f_{k}^{c}=E_{k}f_{k}^{c},  \label{reduced H}
\end{equation}%
i.e., formally $\bar{H}_{c}$\ can lead the same result as the original
Hamiltonian acted with respect to the whole graph, then is referred as the
effective Hamiltonian for central graph. Nevertheless, in general, one
cannot treat Eq. (\ref{reduced H}) as usual since it is hard to obtain the
explicit matrix form of $\bar{H}_{c}$.

\section{Effective Hamiltonian for central graph}

\label{sec_central_graph} It can be seen from Eq. (\ref{H_eff_c}) that, $%
\bar{H}_{c}$ is constructed based on the original subgraph $H_{c}$. It
indicates that the impact of two branch graphs can be projected on the
target graph as additional couplings or on-site potentials. In this paper,
we investigate a graph with each independent branch graph connected to the
central graph $c$\ via a \textit{single} node on the central graph. This is
crucial and our conclusion is available for a graph with arbitrary branches.
In the following we will show that $\bar{H}_{a}$ and $\bar{H}_{b}$ have a
concise form and clear physical meaning.

The connections between the subgraphs are

\begin{eqnarray}
H_{ca} &=&H_{ac}^{\dagger }=-\sum_{j}^{N_{a}}g_{j}^{a}\left\vert
A\right\rangle _{ca}\left\langle j\right\vert ,  \label{H_ca} \\
H_{cb} &=&H_{bc}^{\dagger }=-\sum_{j}^{N_{b}}g_{j}^{b}\left\vert
B\right\rangle _{cb}\left\langle j\right\vert .  \label{H_cb}
\end{eqnarray}%
Note that there is only one branch-root node for each branch, that is the
unique restriction to the graph.

We note from Eq. (\ref{H_ca}) that the elements of $H_{ca}$ and $%
H_{ac}^{\dag }$\ are all zeros except the row connecting to node $A$, i.e.,

\begin{equation}
H_{ca}\left( m,n\right) =\delta _{mA}g_{n}^{a},\text{ }H_{ac}\left(
m,n\right) =\delta _{nA}\left( g_{m}^{a}\right) ^{\ast }.  \label{EL_H_ca}
\end{equation}%
\ Taking $M^{a}=\left( E_{k}-H_{a}\right) ^{-1}$ and assuming its existence
for the considering eigenvalue $E_{k}$, we have

\begin{eqnarray}
\bar{H}_{a}\left( m,n\right) &=&\sum_{j^{\prime
}=1}^{N_{a}}[\sum_{j=1}^{N_{a}}H_{ca}\left( m,j\right) M^{a}\left(
j,j^{\prime }\right) ]H_{ac}\left( j^{\prime },n\right)  \notag \\
&=&\sum_{j^{\prime }=1}^{N_{a}}[\sum_{j=1}^{N_{a}}\delta
_{mA}g_{j}^{a}M^{a}\left( j,j^{\prime }\right) ]\delta _{nA}(g_{j^{\prime
}}^{a})^{\ast }  \notag \\
&=&\delta _{mA}\delta _{nA}\sum_{j,j^{\prime
}=1}^{N_{a}}g_{j}^{a}(g_{j^{\prime }}^{a})^{\ast }M^{a}\left( j,j^{\prime
}\right) .
\end{eqnarray}%
Moreover, from Eqs. (\ref{fk_a}) and (\ref{H_a_eff}) we obtain

\begin{equation}
H_{ca}f_{k}^{a}=\bar{H}_{a}f_{k}^{c}
\end{equation}%
and its explicit form

\begin{equation}
\sum_{j=1}^{N_{a}}g_{j}^{a}f_{k}^{a}\left( j\right) =f_{k}^{c}\left(
A\right) \sum_{j,j^{\prime }=1}^{N_{a}}g_{j}^{a}(g_{j^{\prime }}^{a})^{\ast
}M^{a}\left( j,j^{\prime }\right) .
\end{equation}%
Considering the non-trivial case $f_{k}^{c}\left( A\right) \neq 0$, the
effective Hamiltonian $\bar{H}_{a}$ can be expressed as

\begin{equation}
\bar{H}_{a}\left( m,n\right) =\frac{\delta _{mA}\delta _{nA}}{%
f_{k}^{c}\left( A\right) }\sum_{j=1}^{N_{a}}g_{j}^{a}f_{k}^{a}\left(
j\right) .  \label{H_a_eff_concrete}
\end{equation}%
By a similar procedure we obtain expression for the effective Hamiltonian $%
\bar{H}_{b}$

\begin{equation}
\bar{H}_{b}\left( m,n\right) =\frac{\delta _{mB}\delta _{nB}}{%
f_{k}^{c}\left( B\right) }\sum_{j=1}^{N_{b}}g_{j}^{b}f_{k}^{b}\left(
j\right) .  \label{H_b_eff_concrete}
\end{equation}%
Surprisingly, matrix $\bar{H}_{a}$ ($\bar{H}_{b}$) contains only one nonzero
element $\bar{H}_{a}(A,A)$ ($\bar{H}_{b}(B,B)$), which can be regarded as an
effective on-site potential at the branch-root node $A$ ($B$). Actually,
this is caused by the unique restriction. Then the physics of the projection
Hamiltonian is very clear: original target Hamiltonian with additional
potentials at the joint sites. The effective potential is a weighted
summation of the coupling strength $\{g_{j}^{\gamma =a,b}\}$ and the
corresponding amplitudes $\{f_{k}\left( j\right) \}$. It would be noted that
this conclusion can be generalized into graphs with more independent
branches $d,e,\cdots $.

One can simply classify the branch graph as finite or infinite. For finite
graph without flux, we have $\{g_{j}^{\gamma =a,b,\cdots }\}$ and the
corresponding $\left\{ f_{k}\left( j\right) \right\} $ are all real, then
the effective on-site potentials are real. In contrary, for an infinite
graph, when dealing with the scattering problem, the effective on-site
potentials could be complex.

\section{ILLUSTRATIVE EXAMPLES}

\label{sec_illus} In this section, two typical examples, which consist of
finite and infinite branch graphs, are respectively investigated to
exemplify the formalism developed above.

\subsection{Finite chain}

We first take a finite chain $N$ as an example, with the Hamiltonian in the
form

\begin{equation*}
H_{\text{Chain}}=-J\sum_{i=1}^{N-1}\left( \left\vert i\right\rangle
\left\langle i+1\right\vert +\text{H.c.}\right) .
\end{equation*}%
It is well known that the eigenvalue $E_{k}$\ and the corresponding
eigenvector $f_{k}$ are

\begin{eqnarray}
E_{k} &=&-2J\cos k,  \label{eigenvalue_Chain} \\
f_{k}\left( j\right) &=&\sqrt{\frac{2}{N+1}}\sin \left( kj\right) ,
\label{eigenvector_Chain} \\
k &=&\frac{n\pi }{N+1},n\in \lbrack 1,N].  \notag
\end{eqnarray}%
Now we divide the chain $N$ as the central part $N_{c}$ and two branches $%
N_{a}$, $N_{b}$\ as mentioned above. The two branch-root nodes are located
at the $\left( N_{a}+1\right) $th and $\left( N_{a}+N_{c}\right) $th sites.
From Eqs. (\ref{H_a_eff_concrete}) and (\ref{H_b_eff_concrete}), the
projection Hamiltonian can be obtained as%
\begin{eqnarray}
\mathcal{\bar{H}}_{c} &=&-J\sum_{i=N_{a}+1}^{N_{a}+N_{c}-1}\left( \left\vert
i\right\rangle \left\langle i+1\right\vert +\text{H.c.}\right)
\label{Hbar_C} \\
&&+V_{A}\left\vert N_{a}+1\right\rangle \left\langle N_{a}+1\right\vert
+V_{B}\left\vert N_{a}+N_{c}\right\rangle \left\langle
N_{a}+N_{c}\right\vert ,  \notag
\end{eqnarray}%
where the on-site potentials are%
\begin{eqnarray}
V_{A} &=&-J\frac{\sin \left( kN_{a}\right) }{\sin \left[ k\left(
N_{a}+1\right) \right] },  \label{Chain_Va} \\
V_{B} &=&-J\frac{\sin \left[ k\left( N_{a}+N_{c}+1\right) \right] }{\sin %
\left[ k\left( N_{a}+N_{c}\right) \right] }.  \label{Chain_Vb}
\end{eqnarray}%
In the Appendix \ref{App_chain}, it is shown that $E_{k}$ is always the
eigenvalue of $\mathcal{\bar{H}}_{c}$ and the corresponding eigenvector of $%
\mathcal{\bar{H}}_{c}$ accords with that of $H_{\text{Chain}}$ within the
central chain $c$. It is noted that potential $V_{A}$ ($V_{B}$)\ does not
exists in the case $\sin \left[ k\left( N_{a}+1\right) \right] =0$ ($\sin %
\left[ k\left( N_{a}+N_{c}\right) \right] =0$). Actually, the corresponding
eigenfunction has vanishing amplitude at the node $A$ ($B$), and $E_{k}$ is
also the eigenvalue of the branch Hamiltonian $H_{a}$ ($H_{b}$)
simultaneously. From the viewpoint of the projection theory, the
corresponding inverse matrix $\left( E_{k}-H_{a}\right) ^{-1}$ or $\left(
E_{k}-H_{b}\right) ^{-1}$ does not exist.

Now we look at a concrete example in order to give a sense of the
conclusion. Consider a $15$-site chain with $N_{a}=5$, $N_{c}=4$, and $%
N_{b}=6$. Taking $k=\pi /4$\ as an example, the corresponding eigenvalue and
eigenvector for the entire chain are $E_{\pi /4}=-\sqrt{2}J$, $f_{\pi
/4}\left( j\right) $ $=\sqrt{2}/4\sin \left( j\pi /4\right) $,\ on the
central chain $c$, $f_{\pi /4}^{c}$ $=-(\sqrt{2},1,0,-1)/4$. On the other
hand, from Eqs. (\ref{eigenvalue_Chain}), (\ref{eigenvector_Chain}), (\ref%
{Chain_Va}) and (\ref{Chain_Vb}), we have $V_{A}$ $=-J\sin \left( 5k\right)
/\sin \left( 6k\right) $ $=-\sqrt{2}/2J$ and $V_{B}$ $=-J\sin \left(
10k\right) /\sin \left( 9k\right) $ $=-\sqrt{2}J$. Then the corresponding
effective Hamiltonian is

\begin{equation}
H_{c}^{\text{eff}}=-J\sum_{i=6}^{8}\left( \left\vert i\right\rangle
\left\langle i+1\right\vert +\text{H.c.}\right) -J(\frac{\sqrt{2}}{2}%
\left\vert 6\right\rangle \left\langle 6\right\vert +\sqrt{2}\left\vert
9\right\rangle \left\langle 9\right\vert ),
\end{equation}%
to solve $H_{c}^{\text{eff}}$, we use the formula Eq. (\ref{critical eq1})
derived in the Appendix \ref{App_chain}. It becomes

\begin{equation}
\sin \left( 4\kappa \right) (2\cos \kappa -\frac{3\sqrt{2}}{2})=0,
\end{equation}%
which has the solutions $E_{\kappa }=-2J\cos \kappa $ $=\sqrt{2}J$, $0$, $-%
\sqrt{2}J$, and $-3\sqrt{2}/2J$. The corresponding eigenvector for $%
E_{\kappa }=-\sqrt{2}J$ can be obtained as $(f_{\pi /4}^{\text{eff}%
})^{\dagger }\propto (\sqrt{2},1,0,-1)$, which accords with wavefunction of
whole system within the chain $c$, $f_{\pi /4}^{c}$.

%%%%%%%%%%%%%%%%%%%%%%%%%%%%%%%%%%%%%%%%%%%%%%%%%%%%%%%%%%%%%%%%%%%%%%%%

\begin{figure}[tbp]
\includegraphics[ bb=0 175 577 618, width=6.5 cm, clip]{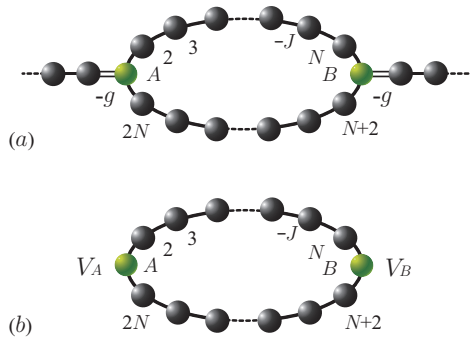}
\caption{(Color online) Schematic illustration of the concrete configuration for a scattering system.
A ring as the scattering center, connects to two semi-infinite chains $L$ and $R$ as waveguides with coupling
$-g$. The wave function within the scattering center for a
scattering state of the whole system is identical to an equal-energy
eigen function of the projection Hamiltonian which is constructed by the
center ring with additional on-site potentials $V_A$ and $V_B$ added at the
joint sites $A$ and $B$.} \label{ring}
\end{figure}

%%%%%%%%%%%%%%%%%%%%%%%%%%%%%%%%%%%%%%%%%%%%%%%%%%%%%%%%%%%%%%%%%%%%%%%%

\subsection{Scattering problem}

In the above example, we can see that all the potentials are real. It was
predicted that the infinite branches could induce the imaginary potentials.
Here we are interested in the scattering solution of an infinite system.
Quantum scattering and transport properties in quantum networks are
important features in quantum information science \cite{Kim,ZhouL,JLTrans}.
Now we consider an exactly solvable but non-trivial system to illustrate the
main idea of this paper.

The graph is constructed by a uniform ring system and two semi-infinite
chains as the input and output leads, which is schemed in Fig. (\ref{ring}).
It is worthy to point that well-established Green function technique \cite%
{JLTrans,Datta,YangAs} can be employed to obtain the reflection and
transmission coefficients for a given incoming plane wave. The corresponding
wave function within the scattering center should be obtained via Bethe
ansatz method. The Hamiltonian can be written as

\begin{eqnarray}
&&H_{s}=\mathcal{H}_{a}+\mathcal{H}_{b}+\mathcal{H}_{c}  \label{H_scat} \\
&&-\sqrt{2}J\left( \left\vert -1\right\rangle _{ac}\left\langle 1\right\vert
+\left\vert 1\right\rangle _{bc}\left\langle N+1\right\vert +\text{H.c.}%
\right) \text{,}  \notag
\end{eqnarray}%
with $\mathcal{H}_{a}$ ($\mathcal{H}_{b}$) represents a uniform input
(output) waveguide as

\begin{eqnarray}
\mathcal{H}_{a} &=&-J\sum_{i=-1}^{-\infty }\left( \left\vert
i-1\right\rangle _{a}\left\langle i\right\vert +\text{H.c.}\right) , \\
\mathcal{H}_{b} &=&-J\sum_{i=1}^{+\infty }\left( \left\vert i\right\rangle
_{b}\left\langle i+1\right\vert +\text{H.c.}\right) ,
\end{eqnarray}%
and the uniform ring as the scattering center is described as

\begin{eqnarray}
\mathcal{H}_{c} &=&-J\sum_{i=1}^{2N}\left( \left\vert i\right\rangle
_{c}\left\langle i+1\right\vert +\text{H.c.}\right) \\
&&-V\left( \left\vert 1\right\rangle _{c}\left\langle 1\right\vert
+\left\vert N+1\right\rangle _{c}\left\langle N+1\right\vert \right) ,
\notag
\end{eqnarray}%
where $\left\vert 2N+1\right\rangle _{c}\equiv \left\vert 1\right\rangle
_{c} $.

There are on-site potentials $V$ at the site $\left\vert 1\right\rangle _{c}$
and $\left\vert N+1\right\rangle _{c}$, which are the two branch-root nodes,
i.e., $\left\vert A\right\rangle _{c}=\left\vert 1\right\rangle _{c}$ and $%
\left\vert B\right\rangle _{c}=\left\vert N+1\right\rangle _{c}$. The
corresponding L\"{o}wdin's projection Hamiltonian depends on the energy $%
E_{k}$ of the incident plane wave as well as the parameter $V$. To be
concise, as an illustrative example, we would like to present the exactly
solvable model, which is helpful to demonstrate our main idea. Therefore, we
will focus on the case: the incident wave has energy $E_{k}=V$ $\in (-2J,2J)$%
. For such an incident plane wave, the scattering wave function can be
obtained by the Bethe ansatz method. The wavefunction has the form,

\begin{eqnarray}
f_{k}^{a}\left( l\right) &=&e^{ik\left( l+1\right) }\text{, }l\in (-\infty
,-1],  \label{fl_a} \\
f_{k}^{c}\left( l\right) &=&e^{ikl}/\sqrt{2}\text{, }l\in \left[ 1,N+1\right]
,  \label{fl_c} \\
f_{k}^{b}\left( l\right) &=&e^{ik\left( l+N+1\right) }\text{, }l\in \lbrack
1,\infty ),  \label{fl_b}
\end{eqnarray}%
where $f_{k}^{c}\left( l\right) \equiv f_{k}^{c}\left( 2N+2-l\right) $. Then
the effective Hamiltonian $\mathcal{\bar{H}}_{a}$, $\mathcal{\bar{H}}_{b}$
can be obtained directly from Eqs. (\ref{H_a_eff_concrete}) and (\ref%
{H_b_eff_concrete}), which have the form

\begin{eqnarray}
\mathcal{\bar{H}}_{a}\left( A,A\right) &=&-\frac{\sqrt{2}Jf_{k}^{a}\left(
-1\right) }{f_{k}^{c}\left( 1\right) }=-2Je^{-ik}, \\
\mathcal{\bar{H}}_{b}\left( B,B\right) &=&-\frac{\sqrt{2}Jf_{k}^{b}\left(
1\right) }{f_{k}^{c}\left( N+1\right) }=-2Je^{ik}.
\end{eqnarray}%
The projection Hamiltonian $\mathcal{\bar{H}}_{c}$ ($\mathcal{\bar{H}}_{c}=%
\mathcal{H}_{c}+\mathcal{\bar{H}}_{a}+\mathcal{\bar{H}}_{b})$ is

\begin{eqnarray}
\mathcal{\bar{H}}_{c} &=&-J\sum_{i=1}^{2N}\left( \left\vert i\right\rangle
_{c}\left\langle i+1\right\vert +\left\vert i+1\right\rangle
_{c}\left\langle i\right\vert \right)  \label{Projection_H_c} \\
&&+2iJ\sin k\left\vert 1\right\rangle _{c}\left\langle 1\right\vert -2iJ\sin
k\left\vert N+1\right\rangle _{c}\left\langle N+1\right\vert .  \notag
\end{eqnarray}

It is a $\mathcal{PT}$ symmetric non-Hermitian Hamiltonian. Since the
seminal discovery by Bender \cite{Bender 98}, it is found that non-Hermitian
Hamiltonian with simultaneous unbroken $\mathcal{PT}$ symmetry has an
entirely real quantum mechanical energy spectrum and has profound
theoretical and methodological implications. In the Appendix \ref{App_scat},
it is shown the spectrum $\{\varepsilon \}$ of $\mathcal{\bar{H}}_{c}$\
consists of a band%
\begin{eqnarray}
\varepsilon _{j} &=&-2J\cos \left( j\pi /N\right) ,  \label{E_i} \\
\text{( }j &\in &\left[ 1,N-1\right] ,\text{2-fold degeneracy)}  \notag
\end{eqnarray}%
and two additional levels%
\begin{equation}
\varepsilon _{\pm }=\pm V.  \label{E+/-}
\end{equation}%
The eigenstates with eigenvalue $\varepsilon _{j}$ can be decomposed into
two kinds: symmetric and anti-symmetric with respect to the spatial
reflection symmetry about the axis along the waveguides. For the scattering
problem, only the symmetric states are involved. It shows that among the
eigenvalues, the eigenvalue $\varepsilon _{+}=V$ from the spectrum $%
\{\varepsilon \}$ matches the energy $E_{k}$ ($E_{k}=V$) of the incident
wave. Moreover, in the end of Appendix \ref{App_scat}, it is shown that the
corresponding eigenvetor for $\varepsilon _{+}$ accords with $f_{k}^{c}$.
Thus it is in agreement with the conclusion of the partitioning technique
that, there always exists a solution of the projection Hamiltonian\ to match
the incident wave energy.

\section{Summary}

\label{sec_conclusion} In summary, we apply the L\"{o}wdin's projection
theory to the specified network, which consists of a central graph and
several branch graphs. It is shown that the effective Hamiltonian on the
central graph can be constructed by adding additional potentials on the
branch-root nodes, which can be expressed as a weighted summation of the
corresponding wavefunction and generates the same result as does the the
original Hamiltonian on the entire graph. It indicates that the impact of
the branch graph to the central one is local and takes the role of the
on-site potential, A finite and an infinite exactly solvable models are
presented to demonstrate our conclusion.

\acknowledgments We acknowledge the support of the CNSF (Grant Nos. 10874091
and 2006CB921205). \appendix

\section{Bethe ansatz solution}

In this Appendix, we will derive the central formula for studying the eigen
problem of the projection Hamiltonian introduced in section \ref{sec_illus}.

\subsection{N-site uniform chain}

\label{App_chain} We consider a uniform chain with potentials at ends. The
projection Hamiltonian is

\begin{equation}
H_{c}^{\text{eff}}=-J\sum_{i=1}^{N_{c}-1}\left( \left\vert i\right\rangle
\left\langle i+1\right\vert +\text{H.c.}\right) +V_{A}\left\vert
1\right\rangle \left\langle 1\right\vert +V_{B}\left\vert N_{c}\right\rangle
\left\langle N_{c}\right\vert  \label{H_eff_Chain_App}
\end{equation}%
where $V_{A}$ and\ $V_{B}$ are defined in Eqs. (\ref{Chain_Va}) and (\ref%
{Chain_Vb}). The Bethe ansatz eigenvector has the form

\begin{equation}
f_{\kappa }=A_{\kappa }e^{i\kappa j}+B_{\kappa }e^{-i\kappa j},j\in \left[
1,N_{c}\right] .  \label{f_chain_App}
\end{equation}%
The Schr\"{o}dinger equation $H_{c}^{\text{eff}}\left\vert f_{\kappa
}\right\rangle =E_{\kappa }\left\vert f_{\kappa }\right\rangle $ can be
written in the explicit form%
\begin{equation}
\begin{array}{r}
V_{A}f_{\kappa }\left( 1\right) -Jf_{\kappa }\left( 2\right) =E_{\kappa
}f_{\kappa }\left( 1\right) , \\
-Jf_{\kappa }\left( j-1\right) -Jf_{\kappa }\left( j+1\right) =E_{\kappa
}f_{\kappa }\left( j\right) , \\
j\in \left[ 2,N_{c}-1\right] , \\
-Jf_{\kappa }\left( N_{c}-1\right) +V_{B}f_{\kappa }\left( N_{c}\right)
=E_{\kappa }f_{\kappa }\left( N_{c}\right) .%
\end{array}
\label{Schrodinger_chain_App}
\end{equation}%
Substituting Eq. (\ref{f_chain_App}) into Eq. (\ref{Schrodinger_chain_App}),
we obtain

\begin{eqnarray}
J^{2}\sin \left[ \kappa \left( N_{c}+1\right) \right] +J\left(
V_{A}+V_{B}\right) \sin \left( \kappa N_{c}\right) &&  \label{crtical eq} \\
+V_{A}V_{B}\sin \left[ \kappa \left( N_{c}-1\right) \right] =0, &&  \notag \\
E_{\kappa }=-2J\cos \kappa . &&  \label{spectrum}
\end{eqnarray}%
Eq. (\ref{crtical eq}) determines the solution of $\kappa $,\ while Eq. (\ref%
{spectrum}) is the corresponding spectrum. Substituting Eqs. (\ref{Chain_Va}%
) and (\ref{Chain_Vb}) into Eq. (\ref{crtical eq}), we have%
\begin{equation}
\begin{array}{r}
\sin k/\left\{ \sin \left[ k\left( N_{a}+1\right) \right] \sin \left[
k\left( N_{a}+N_{c}\right) \right] \right\} \\
\times \left\{ \sin \left[ k\left( N_{c}-1\right) \right] \sin \left( \kappa
N_{c}\right) -\sin \left( kN_{c}\right) \sin \left[ \kappa \left(
N_{c}-1\right) \right] \right\} \\
+2\sin \left( \kappa N_{c}\right) \left( \cos \kappa -\cos k\right) =0%
\end{array}
\label{critical eq1}
\end{equation}%
which seems difficult to solve. However, it can be simply proved by
straightforward algebra that, $\kappa =k$ is a solution for the equation.
Accordingly, $E_{\kappa }=-2J\cos \kappa =-2J\cos k$ is an eigenvalue of the
effective Hamiltonian of Eq. (\ref{H_eff_Chain_App}).\quad Now we try to
find the corresponding eigenvector of $E_{\kappa }$. From Eq. (\ref%
{f_chain_App}), the first equation of Eq. (\ref{Schrodinger_chain_App}) and
the expression of $V_{A}$ Eq. (\ref{Chain_Va}), we obtain

\begin{equation}
\frac{B_{\kappa }}{A_{\kappa }}=-e^{-2ikN_{a}}
\end{equation}%
it indicates%
\begin{equation}
f_{\kappa }\left( j\right) \propto \sin \left[ k\left( N_{a}+j\right) \right]
\end{equation}%
which accords with the eigenfunction Eq. (\ref{eigenvector_Chain}) inside
the central chain $N_{c}$.

\subsection{Uniform ring as a scattering center}

\label{App_scat} The projection Hamiltonian on a uniform ring is $\mathcal{PT%
}$\ symmetric and can be expressed as

\begin{eqnarray}
H_{c}^{\text{eff}} &=&-J\sum_{j=1}^{2N}\left( \left\vert j\right\rangle
\left\langle j+1\right\vert +\text{H.c.}\right)  \label{H_eff_ring_App} \\
&&+2iJ\sin k\left( \left\vert 1\right\rangle \left\langle 1\right\vert
-\left\vert N+1\right\rangle \left\langle N+1\right\vert \right) ,  \notag
\end{eqnarray}%
where $\left\vert j\right\rangle =\left\vert 2N+j\right\rangle $. The parity
operator $\mathcal{P}$ is given by
\begin{equation*}
\mathcal{P}\left\vert j\right\rangle =\pm \left\vert N+2-j\right\rangle
\end{equation*}%
and the time-reversal operator $\mathcal{T}$\ obeys $\mathcal{T}i\mathcal{T}%
^{-1}=-i$. We note that the Hamiltonian $H_{c}^{\text{eff}}$\ also possesses
the mirror symmetry with respect to the axis through the $1$-th and $\left(
N+1\right) $-th sites. This leads to the symmetric and antisymmetric
solutions of the system. The symmetric\ Bethe ansatz eigenfunction $%
f_{\kappa }$ has the form

\begin{equation}
f_{\kappa }\left( j\right) =\left\{
\begin{array}{c}
A_{\kappa }e^{i\kappa j}+B_{\kappa }e^{-i\kappa j}, \\
j\in \left[ 1,N+1\right] \\
A_{\kappa }e^{i\kappa \left( 2N+2-j\right) }+B_{\kappa }e^{-i\kappa \left(
2N+2-j\right) }, \\
j\in \left[ N+2,2N\right]%
\end{array}%
\right. .  \label{BA_f_kappa_ring_App}
\end{equation}%
Substituting the above wave function into the following Schr\"{o}dinger
equation

\begin{align}
2i\sin kf_{\kappa }\left( 1\right) -f_{\kappa }\left( 2\right) -f_{\kappa
}\left( 2N\right) =E_{\kappa }f_{\kappa }\left( 1\right) /J,&  \notag \\
-f_{\kappa }\left( j-1\right) -f_{\kappa }\left( j+1\right) =E_{\kappa
}f_{\kappa }\left( j\right) /J,&  \notag \\
j\in \left[ 2,N\right] \cup \left[ N+2,2N\right] ,&
\label{Ring_Schroding_App} \\
-f_{\kappa }\left( N\right) -f_{\kappa }\left( N+2\right) -2i\sin kf_{\kappa
}\left( N+1\right) &  \notag \\
=E_{\kappa }f_{\kappa }\left( N+1\right) /J,&  \notag
\end{align}%
after simplification, we obtain

\begin{eqnarray}
\left(
\begin{array}{cc}
D_{-} & D_{+} \\
e^{i\kappa N}D_{-} & e^{-i\kappa N}D_{+}%
\end{array}%
\right) \left(
\begin{array}{c}
A_{\kappa }e^{i\kappa } \\
B_{\kappa }e^{-i\kappa }%
\end{array}%
\right) =0, && \\
E_{\kappa }=-2J\cos \kappa , &&
\end{eqnarray}%
where $D_{\pm }=\sin k\pm \sin \kappa $.

The existence of the solution requires

\begin{equation}
\sin \left( \kappa N\right) \left( \sin ^{2}\kappa -\sin ^{2}k\right) =0.
\end{equation}%
The solution is

\begin{eqnarray}
\kappa &=&n\pi /N,n\in \left[ 1,N-1\right] , \\
\kappa &=&k,\pi -k.  \notag
\end{eqnarray}%
the corresponding eigenvalue is Eqs. (\ref{E_i}, \ref{E+/-}).

Obviously, $E_{\kappa }=-2J\cos \kappa =-2J\cos k$ is an eigenvalue of the
effective Hamiltonian Eq. (\ref{H_eff_ring_App}) and the corresponding
eigenvecor is

\begin{equation}
f_{\kappa }\left( j\right) =\left\{
\begin{array}{l}
e^{ikj}\text{, }j\in \left[ 1,N+1\right] \\
e^{ik\left( 2N+2-j\right) }\text{, }j\in \left[ N+2,2N\right]%
\end{array}%
\right. .
\end{equation}%
Therefore, the above eigenfunction $f_{\kappa }$ accords with Eq. (\ref{fl_c}%
).


\begin{thebibliography}{99}
\bibitem{Lindgren} I. Lindgren and J. Morrison, \textit{Atomic Many-Body
Theory} (Springer, Berlin, 1982).

\bibitem{Hubac} I. Huba\v{c} and S. Wilson, \textit{Brillouin-Wigner Methods
for Many-Body Systems} (Springer, New York, 2010).

\bibitem{Feshbach} H. Feshbach, Ann. Phys. \textbf{5}, 357 (1958); Ann.
Phys. \textbf{19}, 287 (1962).

\bibitem{Lowdin} P.-O. L\"{o}wdin, J. Math. Phys. \textbf{3}, 969 (1962).

\bibitem{WXG} X.-G. Wen, \textit{Quantum field theory of many-body systems},
(Oxford University Press, New York, 2004).

\bibitem{Kim} W. Kim, L. Covaci, and F. Marsiglio Phys. Rev. B \textbf{74},
205120 (2006).

\bibitem{FBloch} F. Bloch, Z. Phys. \textbf{52}, 555 (1928).

\bibitem{CZener} C. Zener, Proc. R. Soc. A \textbf{145}, 523 (1934).

\bibitem{Hartmann} T. Hartmann, F. Keck, H. J. Korsch, and S. Mossmann, New
J. Phys. \textbf{6}, 2 (2004).

\bibitem{IBloch} I. Bloch, Nat. Phys. \textbf{1}, 23 (2005).

\bibitem{Zoller} D. Jaksch and P. Zoller, Ann. Phys. \textbf{315}, 52 (2005).

\bibitem{KML} E. Yablonovitch, T. J. Gmitter, and K. M. Leung, Phys. Rev.
Lett. \textbf{67}, 2295 (1991).

\bibitem{UG} U. Gr\"{u}ning, V. Lehmann, and C. M. Engelhardt, Appl. Phys.
Lett. \textbf{66}, 3254 (1995).

\bibitem{ZhouL} L. Zhou, Z. R. Gong, Y. X. Liu, C. P. Sun, and F. Nori,
Phys. Rev. Lett. \textbf{101}, 100501 (2008).

\bibitem{JLTrans} L. Jin and Z. Song, Phys. Rev. A \textbf{81}, 022107
(2010).

\bibitem{Datta} S. Datta, \textit{Electronic Transport in Mesoscopic Systems}
(Cambridge University Press, Cambridge, 1995).

\bibitem{YangAs} S. Yang, Z. Song, and C. P. Sun, arXiv:0912.0324v1.

\bibitem{Bender 98} C. M. Bender and S. Boettcher, Phys. Rev. Lett. \textbf{%
80}, 5243 (1998).
\end{thebibliography}
\end{document}